\documentclass[proof]{WileyASNA-v1}

\articletype{Proceeding}%

\received{XX December 2023}
\revised{XX December 2023}
\accepted{XX December 2023}

\usepackage[greek,english]{babel}
\usepackage{tikz}
\usepackage{graphicx}
\usepackage{amsmath}
\usepackage{tensor}
\usepackage{amssymb}
\usepackage{caption}
\usepackage{mathrsfs}
\usepackage{braket}
\usepackage{slashed}
\usepackage{bbold}
\usepackage{float}
\usepackage{cancel}

\newcommand\normalorder[1]{\mathopen{:}#1\mathclose{:}}

\begin{document}

\title{Effects of a Generalized Uncertainty Principle on the MIT Bag Model Equation of State}

\author[1]{Marcelo Netz-Marzola*}

\author[1,2]{César A. Zen Vasconcellos}

\author[1]{Dimiter Hadjimichef}

\authormark{M. Netz-Marzola \textsc{et al}}

\address[1]{Instituto de F\'isica, Universidade Federal do Rio Grande do Sul (UFRGS), Porto Alegre, Brazil}

\address[2]{International Center for Relativistic Astrophysics Network (ICRANet), Pescara, Italy}

\corres{*M. Netz-Marzola \\ \email{mnmarzola@gmail.com}}

\abstract{The Generalized Uncertainty Principle (GUP) is motivated by the premise that spacetime fluctuations near the Planck scale impose a lower bound on the achievable resolution of distances, leading to a minimum length. Inspired by a semiclassical method that integrates the GUP into the partition function by deforming its phase space, we induce a modification on the thermodynamic quantities of the MIT bag model that we propose serves as an effective semiclassical description of deconfined quark matter in a space with minimal length. We investigate the consequences of this deformation on the zero-temperature limit, revealing a saturation limit for the energy density, pressure and baryon number density and an overall decrease of the thermodynamic quantities which suggests an enhanced stability against gravitational collapse. These findings extend existing research on GUP-deformed Fermi gases. Ultimately, our description introduces effects of quantum gravity in the equations of state for compact stars in a mathematically simple manner, suggesting potential for extension to more complex systems.}


\maketitle

\section{Introduction}

The concept of an indivisible and fundamental physical quantity has permeated human thought across millennia. Democritus, in the 5th century B.C., speculated on the existence of the smallest entities comprising all matter, which he termed \textgreek{ἀτόμους} (atomos), meaning  ``uncuttable'' or ``indivisible.'' This notion of indivisibility laid the philosophical cornerstone for our ongoing quest to understand the underlying structure of nature. 

The current natural and ultimate extension of this ancient consideration lies in the proposition that the fabric of spacetime itself might be composed of indivisible quanta. The emergence of Quantum Field Theory (QFT) further advanced this idea as a minimal length could serve as a solution to the infinities that plagued theoretical calculations. 

Presently, reconciling the quantum field theories that constitute the Standard Model with the principles of General Relativity stands as one of the most profound challenges in theoretical physics. The obstacle is not the inability to quantize gravity itself, but rather that the traditional approach to doing so yields a perturbatively nonrenormalizable theory. At the heart of this issue lies the fact that Newton's constant is a dimensional quantity. This implies the need of an unending sequence of counter-terms in the theory, which ultimately leads to the loss of predictive power.

Remarkably, the presence of a minimal length scale is an inherent feature of all attempts at describing fundamental theories of gravitation. For example, theories of quantum gravity such as String Theory and Quantum Loop Gravity all predict the existence of  a fundamental unit of length \citep{hossenfelder2013minimal,hossenfelder2006note}.

One way of introducing minimal length is through a Generalized Uncertainty Principle (GUP) - as the name suggests, a modification in the Uncertainty Principle. The premise of this argument stems from the assumption that test particles with sufficiently high energies - capable of resolving distances as small as the Planck length - will, due to their gravitational effects, ultimately disrupt the structure of the spacetime they are intended to probe. Consequently, in addition to the well-known quantum uncertainty, there is another uncertainty that emerges from spacetime fluctuations near the Planck scale. If gravity does, in fact, impose a lower bound on the achievable resolution of distances, it may actually serve to regularize quantum field theories rather than rendering them nonrenormalizable.

While black holes, with their exceptional density and gravitational deformation, present an arena where the intertwining of Quantum Mechanics and General Relativity - and consequently the effects of quantum gravity - becomes most pronounced, it is compelling to consider whether compact stars, as the next densest objects known in the universe, might similarly reveal imprints of these fundamental interactions. 

\section{The MIT Bag Model}

\subsection{Thermodynamic quantities of the EoS}

We focus specifically on quark stars, a particular class of compact stars that are composed entirely of quark matter. The conceptual foundations of such stars were established after recognizing that quarks, the elementary fermions that constitute the nucleons, exhibit asymptotic freedom. That is, under conditions of extreme densities or temperatures, quarks effectively become free of interaction, leading to a phase of matter where nucleons lose their individuality and quarks move within a significantly larger region of space \citep{glendenning2012compact}.

Despite the extensive advances in our understanding of the microscopic behavior of quarks, derived mainly from Quantum Chromodynamics (QCD), there are practical limitations when it comes to obtaining the equation of state of dense nuclear matter. This constraint leads us to the MIT Bag Model \citep{chodos1974new}, a much simpler approach based on the principle of asymptotic freedom, which treats quarks as free particles confined within a so-called bag.

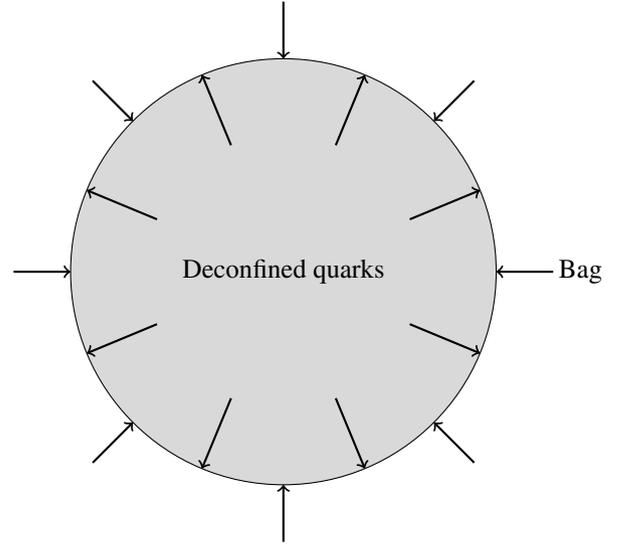
\begin{figure}
    \centering
    \begin{tikzpicture}
        % Draw circle for quark star
        \draw[fill=gray!30] (0,0) circle (2.8cm);
        
        \node at (0,0) {Deconfined quarks};
        
        % Inside arrows
        \foreach \a in {22.5,67.5,...,337.5} {
            \draw[->, thick] (\a:1.8cm) -- (\a:2.8cm);
        }
        
        % Outside arrows
        \foreach \a in {0,45,...,315} {
            \draw[->, thick] (\a:3.55cm) -- (\a:2.8cm);
        }
        
        % Bag constant label
        \node[anchor=west] at (3.5cm,0) {Bag};
    \end{tikzpicture}
    \caption{Representation of the MIT Bag Model.}
    \label{fig:mitbagmodel}
\end{figure}

In the MIT Bag Model, the bag pressure $B$ serves to confine the free quarks within a particular volume. This pressure counteracts the outward pressure exerted by the quarks and ensures the overall stability of the system. As a result, the energy density $\epsilon$ and pressure $\mathscr{P}$ of the quark matter are determined by this confining pressure as well as the kinetic energy of the quarks. In general the thermodynamic quantities of the model can be expressed through:
\begin{align}
\varepsilon&=\frac{1}{(2\pi)^3}\braket{\normalorder{H}}+B\ , \label{a1}\\
\mathscr{P}&=\frac{1}{(2\pi)^3}\frac{\braket{\normalorder{P}}}{3}-B\ ,\label{a2}\\
\rho&=\frac{1}{(2\pi)^3}\frac{\braket{\normalorder{N}}}{3} \ ,\label{a3}
\end{align}
where $\rho$ is the baryon number density and the colons represent normal ordering. $H$, $P$ and $N$ are respectively the Hamiltonian, pressure and number operators. In (\ref{a2}) the factor $3$ divides $\braket{\normalorder{P}}$ since the three identical pressure terms are taken into account, in (\ref{a3}) it divides $\braket{\normalorder{N}}$ because there are three quarks in a Baryon. It should be noted that certain references might choose to implement $B$ within the expectation values, so that it may not always appear explicitly as in equations (\ref{a1}) and (\ref{a2}).

The operators present in (\ref{a1})-(\ref{a3}) are conserved quantities obtained from the Lagrangian spacetime translation and phase shift symmetries of the theory. In the present case, the system of non-interacting quarks is described by the Dirac free-field Lagrangian density  
\begin{equation}
\mathcal{L}= \sum_f \overline{\psi}_f \left( i \gamma^{\mu} \partial_{\mu} - m_f \right) \psi_f \ ,
\end{equation}
which yields \citep{glendenning2012compact}:
\begin{align}
    H&=\sum_f\int_V d\mathbf{x}\, \overline{\psi}_f \left( - i \mathbf{\gamma}\cdot\nabla + m_f \right) \psi_f\ ,\label{a5} \\
    P&=\sum_f\int_V d\mathbf{x} \,\overline{\psi}_f \left( i \gamma^{0} \partial_{0} - m_f \right) \psi_f\ ,\label{a6} \\
    N&=\sum_f \int_V d\mathbf{x}\,  i\psi^\dagger_f \psi_f \ .\label{a7}
\end{align}

The quantization of the fermionic fields in (\ref{a5})-(\ref{a7}) is done through the Fourier expansion \citep{bogoliubov1959introduction,greiner1996field,mandl2010quantum}
\begin{align}
    &\psi_f({x})=\frac{1}{(2\pi)^{3/2}}\int\sum_{r}\left(\frac{m_f}{E_\mathbf{p}}\right)^{(1/2)}\big[c_r(\mathbf{p})u_r(\mathbf{p})e^{-i{p}_\mu{x}^\mu} \nonumber \\ 
    &+d^{\dagger}_r(\mathbf{p})v_r(\mathbf{p})e^{i{p}_\mu{x}^\mu}\big]d\mathbf{p}\ ,\label{fa}\\
    &\overline{\psi}_f ({x})=\frac{1}{(2\pi)^{3/2}}\int\sum_{r}\left(\frac{m_f}{E_\mathbf{p}}\right)^{(1/2)}\big[c^\dagger_r(\mathbf{p})\bar{u}_r(\mathbf{p})e^{i{p}_\mu{x}^\mu}\nonumber \\ 
    &+d_r(\mathbf{p})\bar{v}_r(\mathbf{p})e^{-i{p}_\mu{x}^\mu}\big]d\mathbf{p} \ ,\label{fb}
\end{align}
where  $E_\mathbf{p}=\sqrt{m_f^2+\mathbf{p}^2}$. Using (\ref{fa}) and (\ref{fb}) in (\ref{a5})-(\ref{a7}) and substituting the resulting quantized conserved quantities in the expressions of (\ref{a1})-(\ref{a3}) yields the MIT Bag Model's thermodynamic quantities:
\begin{align}
\varepsilon&=\sum_f\frac{\gamma_f}{2\pi^2}\int_{0}^{\infty} {E_{p}}\Big[n({p},\mu_f)+n({p},-\mu_f)\Big] {p}^2 dp + B\ , \label{a10}\\
\mathscr{P}&=\sum_f\frac{\gamma_f}{6\pi^2}\int_{0}^{\infty} \frac{1}{E_p}\Big[n({p},\mu_f)+n({p},-\mu_f)\Big] {p}^4 dp - B\ ,\label{a11}\\
\rho&=\sum_f\frac{\gamma_f}{6\pi^2}\int_{0}^{\infty} \Big[n({p},\mu_f)-n({p},-\mu_f)\Big] {p}^2 dp\ .\label{a12}
\end{align}
These equations describe the thermodynamic properties of a static compact star comprised of non-interacting deconfined quark matter.

\subsection{EoS in the Zero Temperature Limit}

It is possible to derive an analytic solution for (\ref{a10})-(\ref{a12}) in the particular case of zero temperature. This choice is physically justified if we are to model a compact start in which the nucleons are believed to be dissolved into quarks by the high pressure in the interior of the star: temperatures of neutron stars, shortly after birth, fall into the keV region - which is negligible on the nuclear scale \citep{glendenning2012compact}. For the aforementioned reasons, the analytic solutions we shall find in our present study will all be obtained within this assumption.

In the limit $T\rightarrow0$, the Fermi-Dirac distributions become step functions at energy 
\begin{equation}
\sqrt{m_f^2+p^2}=E_p=\mu_f=\sqrt{m_f^2+p_f^2}\ , \label{a13}
\end{equation}
where $p_f$ is the Fermi momentum. In other words, we have 
\begin{equation}
n(p,\pm\mu_f)\rightarrow\theta(p\mp p_f)\ . \label{a14}
\end{equation}
This simplifies calculations and results in \citep{glendenning2012compact}:
\begin{align}
\varepsilon&=\sum_f\frac{\gamma_f}{2\pi^2}\frac{1}{8}\Big[p_fE_{p_{f}}\left(2p_f^2+m_f^2\right)-m_f^4\omega_f\Big] + B\ ,\label{a15} \\
\mathscr{P}&=\sum_f\frac{\gamma_f}{6\pi^2}\frac{1}{8}\Big[p_fE_{p_{f}}\left(2p_f^2-3m_f^2\right)+3m_f^4\omega_f\Big] - B\ ,\label{a16}\\
\rho&=\sum_f\frac{\gamma_f}{6\pi^2}\frac{1}{3}\,p_f^3\ ,\label{a17}
\end{align}
where $\omega_f=\ln\left(\tfrac{p_f+E_{p_{f}}}{m_f}\right)$ and $E_{p_{f}}=\sqrt{m_f^2+p_f^2}$. 
\section{Kempf Formalism and Deformed Phase Space}\label{sec3}
\subsection{The Kempf GUP}

A Generalized Uncertainty Principle (GUP) is a proposed modification to the standard Heisenberg Uncertainty Principle 
\begin{equation}
\Delta x \Delta p \geq \frac{1}{2}\ .    
\end{equation}
While the Heisenberg Uncertainty Principle establishes a lower bound to the product of the uncertainties in position ($\Delta x$) and momentum ($\Delta p$) of a particle, the GUP introduces additional terms which become especially significant at scales close to the Planck length.

Multiple mental experiments, including a simple Heisenberg microscope in Newtonian gravity \citep{mead1964possible,adler1999gravity} and a more elaborate relativistic version of this argument which considers the measurement of the apparent horizon of a Reissner-Nordstr{\"o}m black hole \citep{maggiore1993generalized} suggest an uncertainty principle of the kind:
\begin{equation}
    \Delta x \Delta p \geq \frac{1}{2}+ \text{const.} (\Delta p)^2 \ . \label{b19}
\end{equation}
Remarkably, (\ref{b19}) corresponds precisely to a known natural consequence of String Theory \citep{konishi1990minimum,chang2011minimal}.

We then follow what we refer to as the Kempf formalism \citep{kempf1995hilbert}, in which we consider a modification of the canonical commutation relation
\begin{equation}
    \left[x,p\right]=i\left(1+\beta p^2\right)\ , \label{b20}
\end{equation}
where $\beta$ is a positive parameter that induces deviations from conventional Quantum Mechanics. Relation (\ref{b20}) then results in a GUP
\begin{equation}
\Delta x \Delta p \geq \frac{1}{2} \left(1 + \beta (\Delta p)^2\right)\ .
\end{equation}
which in turn establishes the minimal length scale
\begin{equation}
    \Delta x \geq x_{min} =\sqrt{\beta}\ .
\end{equation}
For the $n$-dimensional case, however, the deformed Heisenberg algebra is given by the commutation relations \citep{kempf1995hilbert}:
\begin{align}
    &\left[x_i,p_j\right]=i\,\delta_{ij}\left(1+\beta \mathbf{p}^2\right)\ ,\label{b23}\\
    &\left[p_i,p_j\right]=0\ ,\label{b24}\\
    &\left[x_i,x_j\right]=2i\beta\left(x_ip_j-x_jp_i\right) \ ,\label{b25}
\end{align}
where (\ref{b25}) defines a noncommutative geometry.

The relations given in (\ref{b23})-(\ref{b25}) do not break rotational symmetry. Indeed, the generators of rotations may still be written in terms of position and momentum operators as \citep{kempf1995hilbert}
\begin{equation}
    L_{ij}=\frac{1}{1+\beta \mathbf{p}^2}(x_ip_j-x_jp_i)\ .
\end{equation}

These algebra deformations imply profound modifications to the formalism of Quantum Mechanics. While a continuous momentum space is retained as seen from equation (\ref{b24}), the introduction of a noncommutative geometry necessitates the adoption of a quasi-position formalism. In this scenario, even elementary models such as the harmonic oscillator may manifest considerable complexity and notable deviations when the energy scales approach or exceed $\sqrt{\beta}$ \citep{kempf1995hilbert}.

\subsection{The Deformed Poisson Brackets}

From the Hamiltonian point of view of Classical Mechanics, the canonical equations of motion may be represented through Poisson brackets where the position coordinates $x_i$ and the conjugate momenta $p_j$ obey the Poisson algebra, namely
\begin{align}
    &\{x_i,p_j\}=\delta_{ij}\ ,\\
    &\{p_i,p_j\}=0\ ,\\
    &\{x_i,x_j\}=0 \ .
\end{align}

These relationships can be understood physically as enabling the simultaneous measurement of a particle's position and momentum. They imply the exact determination of coordinates in the corresponding phase space, without uncertainty. Transitioning to Quantum Mechanics from this perspective is direct; the classical dynamical variables are replaced with their corresponding Hermitian operators in Hilbert space and Poisson brackets are substituted with Dirac commutators. Consequently, the initial Poisson algebra transforms into the Heisenberg algebra
\begin{align}
    &\left[x_i,p_j\right]=i\,\delta_{ij}\ ,\\
    &\left[p_i,p_j\right]=0\ ,\\
    &\left[x_i,x_j\right]=0 \ ,
\end{align}
which implies the Heisenberg Uncertainty Principle.

A GUP, which induces a deformation to the Heisenberg algebra, should therefore also induce deformation to its classical limit, the Poisson algebra. We consider general deformations to the usual canonical commutation relations, such that
\begin{align}
    &[x_i,p_j]=i\, f_{ij}(x,p) \ \longrightarrow \ \{x_i,p_j\}=f_{ij}(x,p) \ ,\\
    &[p_i,p_j]=i\, h_{ij}(x,p) \ \longrightarrow \ \{p_i,p_j\}=h_{ij}(x,p) \ ,\\
    &[x_i,x_j]=i\, g_{ij}(x,p) \ \longrightarrow \ \{x_i,x_j\}=g_{ij}(x,p) \ ,
\end{align}
 where the deformation functions $f_{ij}$, $g_{ij}$ and $h_{ij}$ are restricted according to the properties of commutators and brackets: bilinearity, the Leibniz rules and the Jacobi identity.

 In the particular case of the Kempf formalism. The commutation relations (\ref{b23})-(\ref{b25}), induce the deformed Poisson brackets:
\begin{align}
    &\{x_i,p_j\}=\delta_{ij} \left(1+\beta p^2\right)\ ,\\
    &\{p_i,p_j\}=0\ ,\\
    &\{x_i,x_j\}=2\beta \left(p_ix_j-p_jx_i\right)\ .
\end{align}

\subsection{Deformation of Differential Volumes}
From the perspective of classical Statistical Mechanics, the coordinates $(x_i,p_i)$ of a particle define a phase space. Since both position and momentum vary with time, the dynamical behavior of the system can be viewed as a continuous trajectory of the phase point in the phase space. In quantum Statistical Mechanics, however, the particle has no well-defined trajectory in the phase space. The Heisenberg Uncertainty Principle, and therefore the Heisenberg algebra, effectively implies a discretization of the phase space in minimal volumes. A modification of the Heisenberg Uncertainty Principle (i.e., a GUP), therefore, deforms such volumes.

In \citep{fityo2008statistical} in particular, the phase space deformation effects are analyzed for the case of a partition function of a quantum system that is then taken to the semiclassical limit. For a non-deformed Heisenberg algebra, we have the transition
\begin{equation}
    Z=\sum_n e^{-E_n/T} \ \longrightarrow \ \ Z=\int e^{-H(X,P)/T}\, {d^{\textit{{N}}}X\,d^{{\textit{N}}}P} \ . \label{b39}
\end{equation}

In the case of a deformed algebra, we know that the expression for the quantum partition function will remain unaltered, since a sum over deformed volumes will have the exact same form as the sum over non-deformed ones. However, this is not the case for a classical partition function. Now we must deal with an integration over phase space that deviates from the original continuous partition function in (\ref{b39}) by a Jacobian factor
\begin{equation}
    J=\frac{\partial(x,p)}{\partial(X,P)} 
\end{equation}
that distorts the phase space, relating the canonical variables of the non-deformed algebra (which we called $X$ and $P$) to the ones of the deformed algebra $x$ and $p$. Namely, we have 
\begin{equation}
    Z=\sum_n e^{-E_n/T} \ \longrightarrow \ \ Z=\int e^{-H(x,p)/T}\, \frac{d^{\textit{N}}x\,d^{\textit{N}}p}{J} \ .
\end{equation}

The Jacobian of the transformation between non-deformed and deformed algebras can be written purely as combinations of deformed Poisson brackets, and, in the particular case of the Kempf deformed algebra of a $2N$-dimensional phase space, we have \citep{fityo2008statistical,chang2002effect}
\begin{equation}
    J=\prod^{N}_{i=1}\,\{x_i,p_i\}=\left(1+\beta p^2\right)^N \ .
\end{equation}
This is an important result that gives us the possibility to calculate (for general deformations) the continuous function without introducing canonically conjugated auxiliary variables. 

In other words, we may effectively incorporate the effects of a GUP into semiclassical $2N$-dimensional systems by applying the following transformation to their non-deformed counterparts:
\begin{equation}
    d^{\textit{{N}}}x\,d^{\textit{{N}}}p \ \longrightarrow \ \frac{d^{\textit{\tiny{N}}}x\,d^{\textit{\tiny{N}}}p}{(1+\beta p^2)^N} \ ,\label{b43}
\end{equation}
which represents a distortion in the differential volumes of phase space and can be shown to be invariant under time evolution from  the Liouville theorem \citep{chang2002effect,chang2011minimal}. The effective formalism described here has been derived equivalently as a deformation of the Planck constant $h$ in \citep{lubo2000thermodynamic, rama2001some}.

Through deformations of the densities of state for ideal non-relativistic and ultra-relativistic gases, \citep{wang2010quantum,vakili2012thermostatistics} derive corrections for the usual thermodynamic relations. Additionally \citep{wang2010quantum}, specifically within the context of compact star configurations, apply such gas models to the Newtonian gravity-pressure balance equation, \citep{mathew2021existence} do the same but applies it to the TOV equations instead, both suggest small corrections to the Chandrasekhar limit.

\section{The GUP-Deformed MIT Bag Model}
We now use the effective Kempf GUP formalism described in the third section to model the effects of Quantum Gravity in the EoS of compact stars. Specifically, we induce a deformation on the thermodynamic quantities of the MIT Bag Model that we propose serves as an effective semiclassical description of deconfined quark matter in a space with minimal length. We subsequently investigate the zero temperature limit, for which we find analytical solutions.

\subsection{Deformed EoS Thermodynamic Quantities}
We know from Section 2 that the general form of the MIT Bag Model given by (\ref{a1})-(\ref{a3}) is dependent on the mean-valued quantities $\braket{\normalorder{H}}$, $\braket{\normalorder{P}}$ and $\braket{\normalorder{N}}$. In order to be explicitly calculated, all of these quantities must be obtained from their respective Lagrangian density symmetries in second quantization. This process involves integrating the expectation values of the QFT expressions over the space occupied by the involved particles (in this case noninteracting fermions) and all their possible momentum states. 

From the previous section, we know that the effective consequence of the deformation of a Heisenberg algebra due to a GUP is a deformation of the differential volumes involved in the integration of quantities in the semiclassical limit. In other words and more intuitively, a change to $\Delta x \Delta p$ has inherent implications for $dx\,dp$.

We notice that the thermodynamic quantities of the EoS are obtained from integrations that satisfy the requirements for the application of the effective GUP formalism. These integrations over position and momentum naturally introduce differential volumes that can be distorted through a Jacobian term to coherently implement the effects of the new underlying algebra. We may therefore develop an effective GUP model of the MIT Bag Model.

We explicitly write the expectation value of the normal-ordered free-field Dirac Hamiltonian in second quantization:
\begin{align}
    &\braket{\normalorder{H}}=\frac{1}{(2\pi)^{3}} \iint \sum_{f,r,r'}\frac{\gamma_f \, m_f}{\left(E_\mathbf{p}E_\mathbf{p'}\right)^{1/2}}\nonumber\\
    &\times\!\Big\langle\!\normalorder{\Big[c^\dagger_r(\mathbf{p})\bar{u}_r(\mathbf{p})e^{i{p}_\mu{x}^\mu}+d_r(\mathbf{p})\bar{v}_r(\mathbf{p})e^{-ip_\mu{x}^\mu}\Big] \left(- i \mathbf{\gamma}\cdot\nabla + m_f \right)\nonumber\\
   &\times\!\left[c_{r'}(\mathbf{p'})u_{r'}(\mathbf{p'})e^{-ip'_\mu{x}^\mu}\!+\!d^{\dagger}_{r'}(\mathbf{p'})v_{r'}(\mathbf{p'})e^{ip'_\mu x^\mu}\right]}\!\Big\rangle\, d^3\!x \, d^3\!p \, d^3\!p' . \label{c44}
\end{align}
Now, the spatial differential element $d^3x$ may be associated with either $d^3p$ or $d^3p'$ (we arbitrarily choose $d^3p$) so that we may induce the volume deformation through transformation (\ref{b43}) as
\begin{equation}
    d^{\tiny{3}}x\ d^{\tiny{3}}p \ \longrightarrow \ \frac{d^{\tiny{{3}}}x\ d^{\tiny{{3}}}p}{(1+\beta p^2)^3} \ .
\end{equation}
We note that there is no ambiguity in our choice here - either association of the spatial element with one of the two momentum ones will yield exactly the same result. Theory guarantees that this must be the case, since $p$ and $p'$ refer to $\psi$ and $\overline{\psi}$, which represent the same particle species and therefore the same set of momentum states. In the end, these integration variables must run over the same momentum space and we must have coherent momentum labels (as is explicitly shown in the second chapter) so that the canonical anticommutation relations of the quantized Dirac field are satisfied and we have a consistent theory. This is also in agreement with the foundational principle of the uncertainty relation, which stipulates that uncertainties are inherently paired with corresponding coordinate-momentum indices.

Therefore, we write the effective GUP version of (\ref{c44}), which we call $\braket{\normalorder{H}}_\beta$, as 
\begin{align}
    &\braket{\normalorder{H}}_\beta=\frac{1}{(2\pi)^{3}} \iint \sum_{f,r,r'}\frac{\gamma_f \, m_f}{\left(E_\mathbf{p}E_\mathbf{p'}\right)^{1/2}}\nonumber\\
    &\times\!\Big\langle\!\normalorder{\!\Big[c^\dagger_r(\mathbf{p})\bar{u}_r(\mathbf{p})e^{i{p}_\mu{x}^\mu}+d_r(\mathbf{p})\bar{v}_r(\mathbf{p})e^{-ip_\mu{x}^\mu}\Big]\left(- i \mathbf{\gamma}\cdot\nabla + m_f \right) \nonumber\\
    &\times\!\left[\!c_{r'}(\mathbf{p'})u_{r'}(\mathbf{p'})e^{-ip'_\mu{x}^\mu}\!+\!d^{\dagger}_{r'}(\mathbf{p'})v_{r'}(\mathbf{p'})e^{ip'_\mu x^\mu}\!\right]\!}\Big\rangle\, \frac{d^3\!x \, d^3\!p\, d^3\!p'}{\left(1\!+\!\beta p^2\right)^3}.
\end{align}

This implies, now in spherical coordinates and with explicit integration limits,
\begin{equation}
    \braket{\normalorder{H}}_\beta\!=\!\!\sum_f\!\gamma_f\!\!\int_{0}^{\infty}\!\! 4\pi\!{E_{p}}\!\Big[\!n({p},\mu_f)\!+\!n({p},-\mu_f)\!\Big]\! {p}^2\!  \frac{dp}{\left(1+\beta p^2\right)^3} .\label{c47}
\end{equation}
Naturally, this same procedure can be applied to $\braket{\normalorder{P}}$ and $\braket{\normalorder{N}}$, yielding: 
\begin{align}
    &\braket{\normalorder{P}}_\beta\!=\!\!\sum_f\!\gamma_f\!\int_{0}^{\infty}\!\! 4\pi\! E_{p}^{-1}\!\Big[\!n({p},\mu_f)\!+\!n({p},-\mu_f)\!\Big]\! {p}^4\! \frac{dp}{\left(1+\beta p^2\right)^3},\label{c48}\\
    &\braket{\normalorder{N}}_\beta\!=\!\!\sum_f\!\gamma_f\!\int_{0}^{\infty}\!\! 4\pi\! \Big[\!n({p},\mu_f)\!-\!n({p},-\mu_f)\!\Big]\! {p}^2\! \frac{dp}{\left(1+\beta p^2\right)^3}.\label{c49}
\end{align}
Having now determined the new deformed quantities (\ref{c47})-(\ref{c49}), we can find the modified MIT Bag Model thermodynamic quantities through the deformed version of relations (\ref{a1})-(\ref{a3}), namely
\begin{align}
\varepsilon_\beta&=\frac{1}{(2\pi)^3}\braket{\normalorder{H}}_\beta+B\ ,\label{c50}\\
\mathscr{P}_\beta&=\frac{1}{(2\pi)^3}\frac{\braket{\normalorder{P}}_\beta}{3}-B\ ,\label{c51}\\
\rho_\beta&=\frac{1}{(2\pi)^3}\frac{\braket{\normalorder{N}}_\beta}{3} \ .\label{c52}
\end{align}
Substituting (\ref{c47})-(\ref{c49}) into (\ref{c50})-(\ref{c52}) respectively, we obtain the effective GUP model for the general MIT Bag Model EoS thermodynamic quantities:
\begin{align}
\varepsilon_\beta&\!=\!\!\sum_f\!\frac{\gamma_f}{2\pi^2}\!\int_{0}^{\infty}\!\! {E_{p}}\!\Big[\!n({p},\mu_f)\!+\!n({p},-\mu_f)\!\Big] {p}^2\!  \frac{dp}{\left(1+\beta p^2\right)^3}\! +\! B\ , \label{c53}\\
\mathscr{P}_\beta&\!=\!\!\sum_f\!\frac{\gamma_f}{6\pi^2}\!\int_{0}^{\infty}\!\! E_{p}^{-1}\!\Big[\!n({p},\mu_f)\!+\!n({p},-\mu_f)\!\Big] {p}^4  \!\frac{dp}{\left(1+\beta p^2\right)^3} \!-\! B\ ,\label{c54}\\
\rho_\beta&\!=\!\!\sum_f\!\frac{\gamma_f}{6\pi^2}\!\int_{0}^{\infty}\! \Big[\!n({p},\mu_f)\!-\!n({p},-\mu_f)\!\Big] {p}^2 \! \frac{dp}{\left(1+\beta p^2\right)^3}\ .\label{c55}
\end{align}
These represent an effective description of a bag of deconfined noninteracting quark matter in a space with a minimal length regulated by the parameter $\beta$ (more specifically, for our choice of units, $x_{min}=\sqrt{\beta}$) induced by a GUP. As expected, the limit $\beta\rightarrow0$ returns us to the conventional quantities (\ref{a10})-(\ref{a12}). There are no further specific restrictions imposed over these quantities other than the ones already incorporated by the usual MIT Bag Model. 

We may therefore insert any physically reasonable set of parameters in (\ref{c53})-(\ref{c55}) and numerically obtain the EoS quantities. However, in order to obtain greater physical insight on the conceptual changes introduced by this model and to be able to later compare it with existing literature, we chose to turn our attention to analytical solutions. More specifically, we shall focus on the zero temperature limit.

\subsection{Deformed EoS in the Zero Temperature Limit}
Now, we investigate the deformed MIT Bag Model in the $T\rightarrow0$ limit. As we have already seen, the Fermi-Dirac distributions become step functions at energy $E_p=\mu_f$. We have thus the effective GUP model equivalent of the quantities found in the second chapter, we now have:
\begin{align}
\varepsilon_\beta&=\sum_f\frac{\gamma_f}{2\pi^2}{\int_{0}^{p_f} \sqrt{m_f^2+p^2}\,{p}^2 \frac{dp}{\left(1+\beta p^2\right)^3}} + B\ , \label{c56}\\
\mathscr{P}_\beta&=\sum_f\frac{\gamma_f}{6\pi^2}{\int_{0}^{p_f} \frac{{p}^4}{\sqrt{m_f^2+p^2}} \,\frac{dp}{\left(1+\beta p^2\right)^3}} - B\ ,\label{c57}\\
\rho_\beta&=\sum_f\frac{\gamma_f}{6\pi^2}{\int_{0}^{p_f} {p}^2 \frac{dp}{\left(1+\beta p^2\right)^3}}\ .\label{c58}
\end{align}
All of these can be integrated to find analytical solutions, namely: 
\begin{align}
&\varepsilon_{\beta}=\sum_f\frac{\gamma_f}{2\pi^2}\frac{m_f}{{8 p_f \left(\beta m_f^2-1\right)^2\left(\beta p_f^2+1\right)^2}}\times\nonumber\\
&\left[\!p_f^2\! \left(\!\beta m_f^2\!-\!1\!\right)\! \sqrt{\frac{(\beta p_f^2\!+\!1)(m_f^2\!+\!p_f^2)}{m_f^2
\left(\beta p_f^2\!+\!1\right)}} \left(m_f^2(\beta p_f^2\!-\!1) \!-\!2 p_f^2\right)\right.\nonumber\\
&\left.+m_f^4 \left(\beta p_f^2\!+\!1\right)^2\! \sqrt{\frac{p_f^2 \left(\beta m_f^2\!-\!1\right)}{m_f^2}} \sin ^{-1}\left(\frac{\sqrt{\frac{p_f^2
\left(\beta m_f^2\!-\!1\right)}{m_f^2}}}{\sqrt{\beta p_f^2\!+\!1}}\right)\!\right]\!+\!B,\label{c59}
\end{align}
\begin{align}
&\mathscr{P}_{\beta}=\sum_f\frac{\gamma_f}{6\pi^2}\frac{-m_f}{8 p_f \left(\beta m_f^2-1\right)^3\left(\beta p_f^2+1\right)^2} \times \nonumber\\
&\left[\!p_f^2\! \left(\beta m_f^2\!-\!1\right)\!\sqrt{\frac{(\beta p_f^2\!+\!1)(m_f^2\!+\!p_f^2)}{m_f^2 \left(\beta p_f^2\!+\!1\right)}} \left(m_f^2(5\beta p_f^2\!+\!3)\!-\!2 p_f^2\right)\right.\nonumber\\
&\left. -3 m_f^4\! \left(\beta p_f^2\!+\!1\right)^2\! \sqrt{\frac{p_f^2
\left(\beta m_f^2\!-\!1\right)}{m_f^2}} \sin ^{-1}\left(\frac{\sqrt{\frac{p_f^2 \left(\beta m_f^2\!-\!1\right)}{m_f^2}}}{\sqrt{\beta p_f^2\!+\!1}}\right)\!\right]\!-\!B,\label{c60}
\end{align}
\begin{equation}
   \rho_\beta = \sum_f\frac{\gamma_f}{6\pi^2} \frac{1}{8\beta^{\frac{3}{2}}}\left[\!\frac{\beta^{\frac{1}{2}}p_f\left(\beta p_f^2-1\right)}{\left({\beta}p_f^2+1\right)^2}+{\tan^{-1}\left(\sqrt{{\beta}}\,p_f\right)}\!\right].\label{c61}
\end{equation}
In the limit $\beta\rightarrow0$, the deformed thermodynamic quantities (\ref{c59})-(\ref{c61}) reduce to the conventional quantities (\ref{a15})-(\ref{a17}), which is required for the consistency of the formalism. More interestingly, if we instead consider the limit where the Fermi momentum $p_f$ goes to infinity, we get
\begin{align}
     &\lim_{p_f\rightarrow\infty}\!\varepsilon_\beta\!=\!\!\sum_f\!\frac{\gamma_f}{16\pi^2}\!\left(\!\frac{\left(1\!-\!\beta m_f^2\!\right)^{-1}\!+\!1}{\beta^2}\!+\!\frac{m_f^4 \sec ^{-1}\!\!\left(\sqrt{\beta} m_f\right)}{\left(\beta m_f^2\!-\!1\right)^{3/2}}\!\right)\!+\!B,\nonumber
\\
    &\lim_{p_f\rightarrow\infty}\!\mathscr{P}_{\beta}\!=\!\!\sum_f\!\frac{\gamma_f}{48\pi^2}\!\left(\!\frac{2\!-\!5 \beta m_f^2}{\beta^2 \!\left(\beta m_f^2\!-\!1\right)^2}\!+\!\frac{3 m_f^4 \sec ^{-1}\!\!\left(\sqrt{\beta} m_f\right)}{\left(\beta m_f^2\!-\!1\right)^{5/2}}\!\right)\!-\!B,\nonumber
\\
    &\lim_{p_f\rightarrow\infty}\!{\rho_\beta}\!=\!\sum_f\frac{\gamma_f}{48\pi^2}\frac{\pi}{2}\,\beta^{-3/2}\ .\label{c62}
\end{align}
What the limits (\ref{c62}) show us is that the energy density, pressure and baryon number density now converge to a maximum saturation value (whereas it would diverge in the conventional case) which scales inversely with respect to the minimal length of the theory. Effectively, the GUP introduces an asymptotic cutoff to the possible momentum configurations.

\begin{figure}[h]
    \centering
    \includegraphics[width=\linewidth]{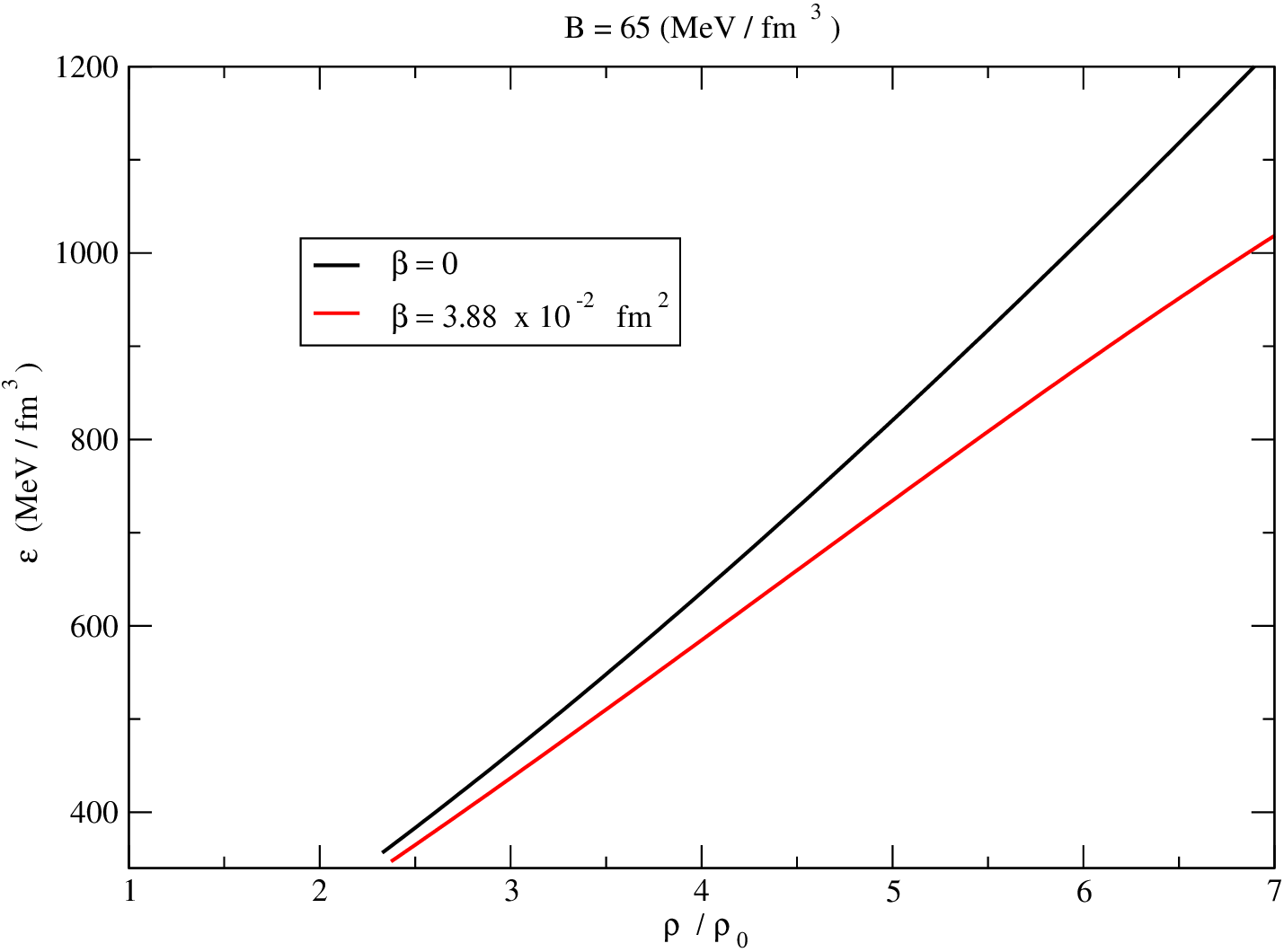}
    \caption{Energy density (we represent $\varepsilon_\beta$ simply as $\varepsilon$) vs. ratio of the baryonic number density (we denote $\rho_\beta$ by $\rho$) and the saturation density ($\rho_0= 0.153\, \text{fm}^{-3}$).}\label{fig2}
\end{figure}

\begin{figure}[h]
    \centering
    \includegraphics[width=\linewidth]{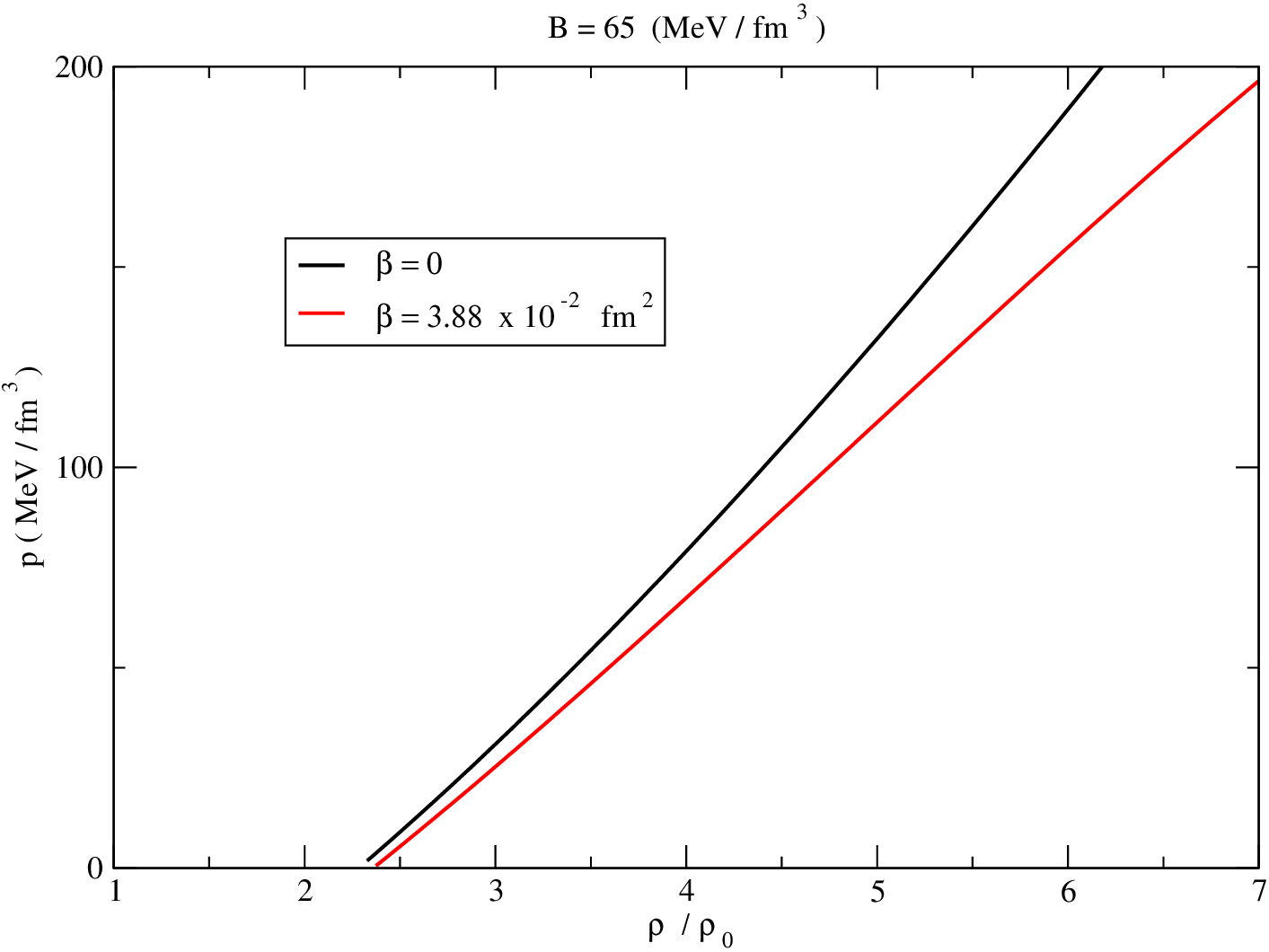}
    \caption{Pressure (we represent $\mathscr{P}_\beta$ simply as $p$) vs. ratio of the baryonic number density ($\rho$) and the saturation density ($\rho_0= 0.153\, \text{fm}^{-3}$).}\label{fig3}
\end{figure}

\begin{figure}[h]
    \centering
    \includegraphics[width=\linewidth]{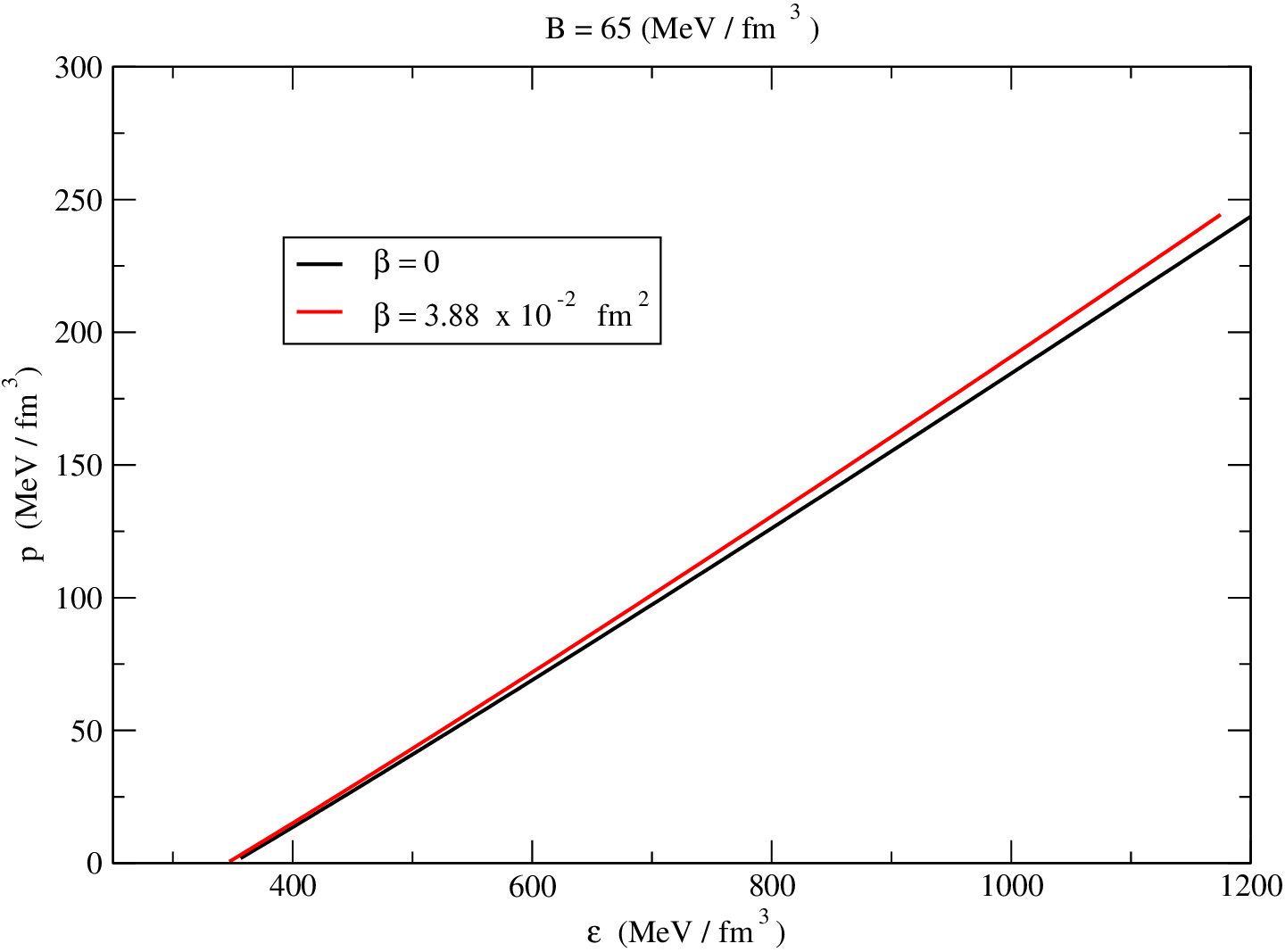}
    \caption{Pressure ($p$) vs. energy density ($\varepsilon$).}\label{fig4}
\end{figure}

Now, knowing that we expect $\beta p^2$ to be much smaller than $1$, we may for the sake simplicity in computations approximate relations (\ref{c56})-(\ref{c58}) up to the first order in $\beta$. Namely, we find
\begin{align}
&\varepsilon_\beta\!=\!\varepsilon_0\!+\!\!\sum_f\!\!\frac{\beta\gamma_f}{32\pi^2}\!\left[\!p_fE_{p_{f}}\!\left(\!-8p_f^4-2p_f^2m_f^2+3m_f^4\!\right)\!-\!3m_f^6\omega_f\!\right],\nonumber \\
&\mathscr{P}_\beta\!=\!\mathscr{P}_0\!+\!\!\sum_f\!\!\frac{\beta\gamma_f}{96\pi^2}\!\left[\!p_fE_{p_{f}}\!\!\left(\!-8p_f^4\!+\!10p_f^2m_f^2\!-\!15m_f^4\!\right)\!\!+\!\!15m_f^6\omega_f\!\right] ,\nonumber\\
&\rho_\beta\!=\!\rho_0\! -\!\! \sum_f\!\!\frac{\beta\gamma_f}{6\pi^2}\frac{3}{5}p_f^5 ,\label{c63}
\end{align}
where $\varepsilon_0$, $\mathscr{P}_0$ and $\rho_0$ are respectively the conventional MIT Bag Model quantities (\ref{a15}), (\ref{a16}) and (\ref{a17}). Taking the limit $\beta\rightarrow0$ returns us to these undeformed values. We may then use the approximate analytical solutions (\ref{c63}) to (within reason, given the limitations of a first-order approximation) analyze the behavior of the thermodynamic quantities of the GUP-deformed MIT Bag Model. The results we obtained for this case are summarized in Figures \ref{fig2} to \ref{fig4} for the order of an upper observational bound of $\beta$ \citep{faruque2014upper}.

From Figures \ref{fig2} and \ref{fig3} we notice that the introduction of the GUP has the effect of, compared to the conventional MIT Bag Model ($\beta=0$), reducing the values of both the energy density and pressure for the same baryon number density. 

We argue that the reduction in the energy density values can be physically justified by the fact that the GUP, as we have noted before, essentially restricts the available momentum states for quarks within the bag. This means that extremely high-momentum (short-wavelength) excitations, which are commonplace in the standard model, get suppressed in the presence of the GUP. This suppression leads to fewer particle excitations in the bag, thereby reducing the overall energy density. 

On the other hand, fewer excitations also mean that the quarks exert less outward pressure on the bag boundaries, thus explaining the decreased pressure. The reduction in pressure, given the same baryon density, would initially suggest that introducing a GUP effectively softens the equation of state in relation to the baryon number density.

However, the relative reductions in energy density occur at a higher rate than those in pressure. This results in the behavior observed in Figure \ref{fig4}, where we see that, for a given energy density, the pressure actually increases compared to the standard MIT Bag Model. This in turn suggests a stiffer EoS, which provides greater resistance to gravitational collapse, meaning we could theoretically find higher mass limits for such compact objects.

In all of the EoS plots, we notice that the effects of the GUP are progressively intensified in higher energy regimes. This is in good accordance with the foundational idea behind the GUP that quantum gravitational effects become more pronounced on higher energy scales. Mathematically, this is due to the momentum dependence of the deformation imposed on the Heisenberg algebra. 

It is also worth noting that in both Figure \ref{fig2} and Figure \ref{fig3}, in addition to the angular divergences between the old and new relations, there is also a slight shift in the lowest value for the baryon number density for each. This is expected from (\ref{c63}).

We conclude this section by observing that, if we reduce our MIT Bag Model to the degenerate Fermi gas model (set $B=0$ in (\ref{c56})-(\ref{c58}) and considering a single particle flavor, for example), and apply the ultra-relativistic limit approximation, our expressions for this constrained version of the EoS have the same form as the ones found for the GUP-modeled degenerate ultra-relativistic Fermi gas in the literature \citep{wang2010quantum,vakili2012thermostatistics,mathew2021existence}. This demonstrates the coherence of our model with previous studies and, more importantly, that we have generalized previous findings. Relations (\ref{c56})-(\ref{c58}) alone are generalizations of the ones already existent in the literature, and, as such, so are the considerably less constrained relations (\ref{c53})-(\ref{c55}).

\section{Final Remarks}\label{sec5}
In this work, we have conducted an investigation into a modification of the equation of state (EoS) for compact stars under a framework that introduces a minimal length scale: the effective Kempf model for the Generalized Uncertainty Principle (GUP). 

The effective GUP formalism was for the first time applied to the MIT Bag Model, resulting in a modified EoS that describes the behavior of noninteracting deconfined quark matter in the presence of a minimal length scale and consistently reduces to the conventional theory when the GUP parameter $\beta\rightarrow0$.

We subsequently derived analytical solutions in the zero temperature limit, for which the thermodynamic quantities presented an overall scaling decrease in relation to the momentum variable. Notably, we have found that as the Fermi momentum goes to infinity, the energy density, pressure and baryon number density converge to a maximum saturation limit. These results are in agreement with the notion that the minimal length scale induced by the GUP imposes an asymptotic cutoff to the possible momentum configurations. Our findings extend existing research on GUP-deformed Fermi gases. 

In particular, Figure \ref{fig4} suggests a stiffer EoS. This provides greater resistance to gravitational collapse, meaning we could theoretically find higher mass limits and radius for such compact objects. Naturally, the subsequent step in this study is investigating the consequences of the modified EoS in the Tolman-Oppenheimer-Volkoff (TOV) equations, this has been done and a paper with such results is currently underway. Additionally, a path towards an even more generalized GUP framework capable of integrating a variety of particles and interactions has been devised and is being prepared for publication. 

We conclude this work by noting that the model studied provides a mathematically simple representation of the potential influence of quantum gravity on compact stellar structures.  We aspire to build upon this work, advancing these models into more complex realms (particularly those involving temperature effects) where quantum gravitational influences may be further amplified. 

\section*{Acknowledgments}
This work was partially funded by CAPES.

\subsection*{Conflict of interest}

The authors declare no potential conflict of interests.

\nocite{*}
\bibliography{Wiley-ASNA}%

\end{document}